# Approaching Waterfilling Capacity of Parallel Channels by Higher Order Modulation and Probabilistic Amplitude Shaping

Fabian Steiner, Patrick Schulte
Institute for Communications Engineering
Technical University of Munich
Email: {fabian.steiner, patrick.schulte}@tum.de

Georg Böcherer
Mathematical and Algorithmic Sciences Lab
Huawei Technologies France
georg.boecherer@ieee.org

*Abstract*—Parallel, additive white Gaussian noise (AWGN) channels with an average sum power constraint are considered. It is shown how the waterfilling Shannon capacity can be approached by higher order modulation and probabilistic amplitude shaping (PAS). This is achieved by a new distribution matching approach called product distribution matching (PDM). The asymptotic performance of PDM is analyzed by achievable rates. A heuristic for optimizing the input distribution is proposed, which enables signaling at a target spectral efficiency with a fixed-rate forward error correction (FEC) code, while the optimal power allocation is ensured by mercury-waterfilling and a simple bit-loading strategy. Finite blocklength simulation results with 5G low-density parity-check codes show power savings of around 1 dB compared to a conventional scheme with uniform input distributions.

## I. INTRODUCTION

Higher-order modulation is indispensable in mobile, satellite, cable, and fiber-optic communication to achieve the high spectral efficiency (SE) required for data applications.

Transceivers must be flexible, i.e., they should support different SEs so they can adapt to the link quality at hand and deliver the best possible connectivity. Conventional coded modulation uses uniform distributions on the constellation points. This has two disadvantages. First, uniform distributions suffer a power inefficiency of up to 1.53 dB [1]. Second, flexibility requires sophisticated modulation and coding approaches, e.g., supporting a large number of modcods (combinations of modulation formats and channel codes such as in DVB-S2X), puncturing schemes (e.g., for Turbo codes in LTE [2, Ch. 10]) or rate-compatible low-density parity-check (LDPC) codes in 5G [3].

One approach to circumvent this deficiency is geometric shaping (GS) [4], [5] which uses constellations with non-equidistant signal points. While improved power efficiency was observed, the problem of flexibility remains. A second approach is probabilistic shaping (PS) [6, p. 208], [7]–[9] that uses equidistant signal points with a non-uniform distribution. For an overview of PS schemes, see [10, Sec. II] and references therein. Recently, we proposed probabilistic amplitude shaping (PAS) [10], a PS architecture that concatenates a distribution matcher (DM) as a shaping device with forward error correction (FEC) for single carrier transmission, see Fig. 1a. PAS achieves the optimal power efficiency and enables flexible SE adaption with a small number of modcods. The benefits of PAS for fiber-optic communication were recently showcased in a transoceanic transmission [11] and future optical modems will implement PAS [12, Sec. V-A].

In many practical settings, the data link is well modeled by a set of non-interacting parallel channels. Examples include multi-carrier transmission such as orthogonal frequency division multiplexing (OFDM), discrete multitone (DMT), and multi-antenna transceivers when the singular value decomposition (SVD) of the channel matrix is used to orthogonalize the system.

In this work, we propose a novel DM architecture called product distribution matching (PDM), which internally uses a collection of parallel DMs with smaller output alphabets to synthesize the desired distribution as product distribution. A preferable implementation uses binary output alphabets for the component DMs. This approach simplifies the implementation significantly and allows for parallelization to achieve high throughput.

We further show how PDM enables PS for multi-carrier transmission by sharing the component DMs for lower bit-levels among different sub-carriers. We provide a representative example where PDM is about 1 dB more power efficient than uniform signaling and operates close to the waterfilling limit [13, Sec. 5.4.6]. The DM implementation used in our simulation is available to the public as a web service at [14].

This work is structured as follows. Sec. II reviews DM and PAS and states achievable rate expressions for system design. In Sec. III, we introduce the PDM architecture and present simulation results for 64-ASK. Sec. IV shows how PDM enables the operation of PAS close to the waterfilling limit of parallel channels. We conclude in Sec. V.

## II. PRELIMINARIES

### A. Distribution Matching (DM)

DMs transform a sequence of uniformly distributed input bits into an output sequence of symbols from an alphabet $\mathcal{A}$ with a desired distribution. In the following, we use the compact notation $x^n = x_1 x_2 \cdots x_n$ for row vectors $(x_1, x_2, \ldots, x_n)$. A fixed-to-fixed length DM maps $k$ input bits $d^k = d_1 d_2 \ldots d_k$ to $n$ output symbols $a^n = a_1 a_2 \ldots a_n = \operatorname{dm}(d^k)$, where $d_i \in \{0,1\}, i = 1, \ldots, k$ and $a_i \in \mathcal{A}, i = 1, \ldots, n$. The mapping $\operatorname{dm}(\cdot)$ is invertible, i.e., $d^k$ can be recovered from $a^n$ by applying the inverse mapping $\operatorname{dm}^{-1}(\cdot)$. Fixed-to-fixed length DMs can be implemented by the constant composition distribution matching (CCDM) [15], for binary output alphabets see also [16]. A DM is specified by the following parameters.

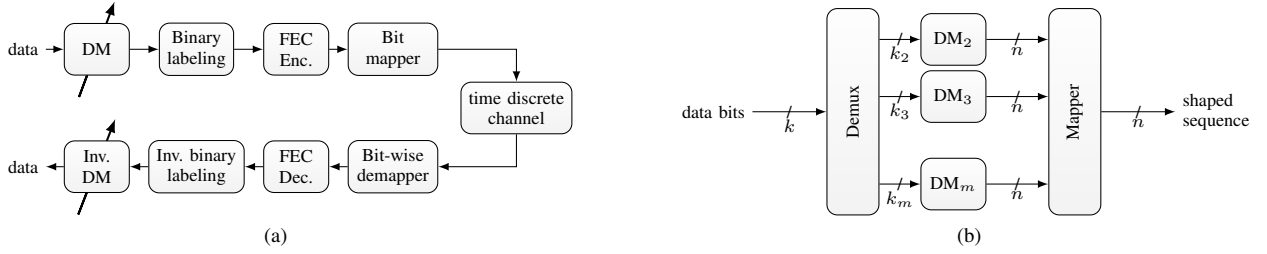

Fig. 1. (a) System model of PAS. The shaping device DM is concatenated in reverse with the FEC device. (b) The DM implementation proposed in this work: Product Distribution Matching (PDM) for $2^m$-ASK. $k$ binary data bits are demultiplexed into $m-1$ parallel blocks of sizes $k_2$ to $k_m$. Parallel binary component DMs output $m$ shaped sequences of length $n$. A bit-mapper recombines the $m-1$ sequences and outputs one shaped sequence of length $n$ that may be used as the amplitudes of the transmit symbols. The employed bit-mapping can be chosen independently from the one used at the receiver side for BMD.

- The rate is
$$R_{\text{dm}} = \frac{k}{n} \quad \left[\frac{\text{bits}}{\text{output symbol}}\right]. \quad (1)$$

- The output distribution is
$$P_A(a) = \frac{\sum_{d^k \in \{0,1\}^k} \bar{P}_{\operatorname{dm}(d^k)}(a)}{2^k}, \quad a \in \mathcal{A} \quad (2)$$

where $\bar{P}_{\operatorname{dm}(d^k)} = \bar{P}_{a^n}$ is the empirical distribution of the sequence $a^n$, i.e.,
$$\bar{P}_{a^n}(a) = \frac{|\{i : a_i = a\}|}{n}, \quad a \in \mathcal{A}. \quad (3)$$

The reference [15, Sec. III.B] shows that $R_{\text{dm}}$ approaches the entropy $\text{H}(A)$ of $P_A$, for large $n$. We take the actual DM rate $k/n$ into account for all finite length numerical results.

### B. Channel Model and Achievable Rates

We consider transmission over the time-discrete additive white Gaussian noise (AWGN) channel
$$Y = X + Z \quad (4)$$

where the channel input $X = \Delta \tilde{X}$ and $\Delta > 0$ is a constellation scaling and $\tilde{X}$ comes from a normalized $M = 2^m$-amplitude shift keying (ASK) constellation
$$\mathcal{X} = \{\pm 1, \pm 3, \ldots, \pm(2^m - 1)\}. \quad (5)$$

The noise $Z$ is zero-mean Gaussian with unit variance, i.e., $Z \sim \mathcal{N}(0,1)$. The resulting signal-to-noise ratio (SNR) is $\text{E}[X^2]/\text{E}[Z^2]$. The mutual information maximizing distribution under an average power constraint is a zero mean Gaussian input $X$ with variance SNR, and the capacity is
$$\text{C}_{\text{AWGN}}(\text{SNR}) = \frac{1}{2} \log_2(1 + \text{SNR}). \quad (6)$$

TABLE I
TWO LABELS FOR 8-ASK. THE AMPLITUDE LABEL OF NBBC IS NBC AND THE AMPLITUDE LABEL OF BRGC IS ALSO BRGC.

|      | -7  | -5  | -3  | -1  | 1   | 3   | 5   | 7   |
|------|-----|-----|-----|-----|-----|-----|-----|-----|
| BRGC | 000 | 001 | 011 | 010 | 110 | 111 | 101 | 100 |
| NBBC | 000 | 001 | 010 | 011 | 111 | 110 | 101 | 100 |

### C. PAS Transmitter

The PAS architecture implements probabilistically shaped ASK modulation [10, Sec. IV.]). It leverages the symmetry of the capacity achieving distribution $P_X$ for the AWGN channel. This allows a factorization of the input distribution into an amplitude and sign part as $P_X(x) = P_A(|x|) \cdot P_S(\operatorname{sign}(x))$, where $P_A$ is non-uniform on $\mathcal{A} = \{|x|, x \in \mathcal{X}\}$ and $S$ is uniform on $\{-1, +1\}$. A DM maps $k$ data bits to $n$ amplitudes $A^n$, which are represented by $n(m-1)$ amplitude bits and associated with bit-levels $b_2 b_3 \cdots b_m$. The amplitude bits and $\gamma n$ additional data bits are multiplied with the parity generating part $\boldsymbol{P}$ of a systematic generator matrix $[\boldsymbol{I}|\boldsymbol{P}]$ to generate $(1-\gamma)n$ parity bits. The parity bits and the additional data bits are mapped to $n$ signs $S^n$ and associated with bit-level $b_1$. The signs are multiplied symbol-wise with the amplitudes $A^n$. The FEC code instantiated by $\boldsymbol{P}$ has rate
$$R_{\text{c}} = \frac{n(m-1) + \gamma n}{mn} = \frac{m-1+\gamma}{m} \quad (7)$$

and the fraction of signs used for data bits is
$$\gamma = 1 - (1 - R_{\text{c}})m. \quad (8)$$

PAS requires $0 \leq \gamma \leq 1$. The transmission rate of PAS is the number of data bits per ASK symbol given by
$$R_{\text{tx}} = \frac{k}{n} + \gamma. \quad (9)$$

### D. Achievable Rates for PAS

In [17], it is shown that an achievable rate for PAS is
$$R_{\text{a}} = \left[\text{H}(X) - \text{E}\left[-\log_2\left(\frac{q(X,Y)}{\sum_{x \in \mathcal{X}} q(x,Y)}\right)\right]\right]^+ \quad (10)$$

where $[\cdot]^+ = \max(0, \cdot)$. The expression $q(x, y)$ is a non-negative memoryless metric on $\mathcal{X} \times \mathbb{R}$ to determine an estimate $\hat{x}^n$ of the sent symbol sequence via

$$\hat{x}^n = \underset{x^n \in \mathcal{C}}{\operatorname{argmax}} \prod_{i=1}^{n} q(x_i, y_i) \quad (11)$$

where $\mathcal{C}$ is the set of codewords. To use binary FEC codes, we introduce a labeling function that maps a constellation point $x \in \mathcal{X}$ to an $m$-bit binary label, i.e., $\chi_{\text{fec}} : \mathcal{X} \to \{0, 1\}^m$ and $\chi_{\text{fec}}(x) = b_1^{\text{fec}} b_2^{\text{fec}} \ldots b_m^{\text{fec}} = \boldsymbol{b}^{\text{fec}}$. Its inverse is $\chi_{\text{fec}}^{-1} : \{0, 1\}^m \to \mathcal{X}$. The BMD metric is

$$q(x, y) = \tilde{q}(\chi_{\text{fec}}(x), y) = \tilde{q}(\boldsymbol{b}^{\text{fec}}, y) = \prod_{i=1}^{m} P_{B_i^{\text{fec}}|Y}(b_i^{\text{fec}}|y) \quad (12)$$

and the achievable rate (10) becomes

$$R_{\text{BMD}}(\text{SNR}; P_X) = \left[ \text{H}(X) - \sum_{i=1}^{m} \text{H}(B_i^{\text{fec}}|Y) \right]^+. \quad (13)$$

In the following, we use a binary reflected Gray code (BRGC) [18] for $\chi_{\text{fec}}(\cdot)$, see Table I.

## III. PRODUCT DISTRIBUTION MATCHING (PDM)

### A. Concept of PDM

Suppose for some amplitude label $b_2^{\text{dm}} \cdots b_m^{\text{dm}}$ and the corresponding signal point label $\boldsymbol{b}^{\text{dm}} = b_1^{\text{dm}} b_2^{\text{dm}} \cdots b_m^{\text{dm}}$ we have

$$P_{\boldsymbol{B}^{\text{dm}}}(\boldsymbol{b}^{\text{dm}}) = \prod_{i=1}^{m} P_{B_i^{\text{dm}}}(b_i^{\text{dm}}) = 0.5 \cdot \prod_{i=2}^{m} P_{B_i^{\text{dm}}}(b_i^{\text{dm}}) \quad (14)$$

where $P_{B_1^{\text{dm}}}(b) = P_{B_1}(b) = 0.5, b \in \{0, 1\}$ because of the required symmetry. The bits of the label $\boldsymbol{b}^{\text{dm}}$ are statistically independent. We can construct a distribution on the signal points by choosing binary distributions $P_{B_i^{\text{dm}}}, i = 2, \ldots, m$ and a bit mapper $\chi_{\text{dm}} : \mathcal{X} \to \{0, 1\}^m$. PDM can efficiently generate the product distributions of (14) and the procedure is displayed in Fig. 1b. $k$ binary data bits are demultiplexed into $m - 1$ parallel blocks of lengths $k_2$ to $k_m$. The $m - 1$ parallel *binary* DMs output $m - 1$ shaped binary sequences of length $n$. A bit mapper recombines the $m - 1$ sequences and outputs one shaped amplitude sequence of length $n$. Hence, PDM provides the same interface as the DM in the original work of [10], which uses one $M/2$-ary DM for a $M$-ASK constellation. The use of binary output DMs greatly reduces the complexity of the underlying matching algorithm, e.g., of arithmetic coding used for CCDM. The rate $R_{\text{dm}}$ for PDM is

$$R_{\text{dm}} = \frac{k_2 + k_3 + \cdots + k_m}{n}. \quad (15)$$

and approaches $\sum_{i=2}^{m} \text{H}(B_i^{\text{dm}})$ for large $n$. An achievable rate asymptotically in $n$ is

$$R_{\text{BMD}}^{\Pi} = \left[ \sum_{i=1}^{m} \text{H}(B_i^{\text{dm}}) - \sum_{i=1}^{m} \text{H}(B_i^{\text{fec}}|Y) \right]^+. \quad (16)$$

Note that the labeling functions $\chi_{\text{dm}}(x)$ and $\chi_{\text{fec}}(x)$ for the DM and FEC may be different for the same signal point. We choose the natural based binary code (NBBC) for $\chi_{\text{dm}}$ and the BRGC for $\chi_{\text{fec}}$ (see Table I) and optimize (16) over the binary distributions $P_{B_i^{\text{dm}}}, i = 2, \ldots, m$ (recall that the sign distribution $P_{B_1}$ is uniform) and the constellation scaling $\Delta$.

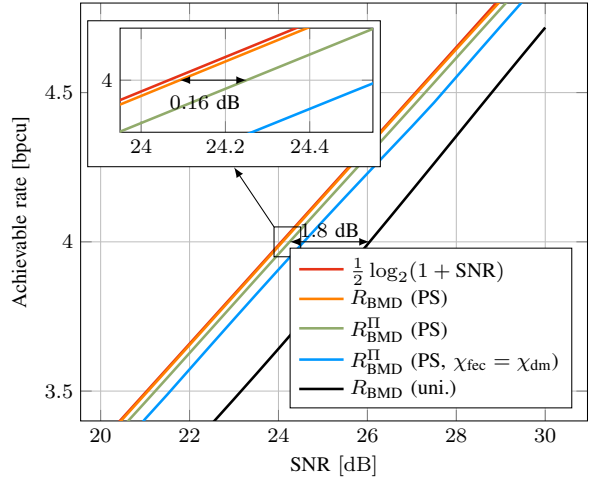

Fig. 2. Achievable rates for 64-ASK and different BMD schemes.

*Remark.* The information-theoretic work [19] only considered the case when $\chi_{\text{fec}} = \chi_{\text{dm}}$, in which case (16) becomes $\sum_{i=1}^{m} \text{I}(B_i^{\text{fec}}; Y)$ which is the so-called bit-interleaved coded modulation (BICM) capacity. We also note that imposing a uniform distribution on $P_X$ implies a uniform distribution on the bit-level probabilities causing them to be statistically independent. In this case, $R_{\text{BMD}}$ also reduces to $R_{\text{BMD}}^{\Pi}$, irrespectively of the employed mapping functions for the DM and FEC.

### B. Achievable Rate Comparisons

In Fig. 2, we display the achievable rates for 64-ASK and different DM schemes. We observe that the product constraint (14) in combination with the different labelings $\chi_{\text{fec}}$ and $\chi_{\text{dm}}$ leads to a performance loss of only 0.16 dB compared to $R_{\text{BMD}}$ with a 32-ary DM at an SE of 4 bpcu (bits per channel use). At the same time, the energy efficiency is improved by 1.8 dB over uniform $R_{\text{BMD}}$. Note that the input distribution has been optimized for the shaped cases of $R_{\text{BMD}}$ and $R_{\text{BMD}}^{\Pi}$ for each SNR.

In Table II, we assess the different DM implementations by their asymptotic achievable rates. We use 64-ASK, a DM amplitude distribution with $R_{\text{dm}} = 4.1$ bits and $\gamma = 0.4$, yielding an SE of $R_{\text{tx}} = 4.5$ bpcu. We employ a 32-ary DM as a reference. The performance of this system is compared to a PDM setup with 1 ($B_2^{\text{dm}}$), 2 ($B_2^{\text{dm}}, B_3^{\text{dm}}$), 3 ($B_2^{\text{dm}}, B_3^{\text{dm}}, B_4^{\text{dm}}$), 4 ($B_2^{\text{dm}}, B_3^{\text{dm}}, B_4^{\text{dm}}, B_5^{\text{dm}}$) and 5 ($B_2^{\text{dm}}, B_3^{\text{dm}}, B_4^{\text{dm}}, B_5^{\text{dm}}, B_6^{\text{dm}}$) individually shaped bit-levels and corresponding binary DMs. The input distributions for PDM have been chosen such that the DMs meet the specified rate using the heuristic

$$\min_{P_{B_2^{\text{dm}}}, \ldots, P_{B_m^{\text{dm}}}} \text{E}\left[X^2\right] \quad \text{s.t.} \quad \sum_{i=2}^{m} \text{H}(B_i^{\text{dm}}) = R_{\text{dm}}. \quad (17)$$

TABLE II
REQUIRED SNRs FOR DIFFERENT DM CONFIGURATIONS AND A TARGET SE OF 4.5 BPCU. (CAPACITY: 27.08 DB)

| DM configuration | Required SNR [dB] | SNR gap to capacity [dB] |
| --- | --- | --- |
| 32-ary DM | 27.13 | 0.05 |
| PDM 1 bit shaped | 28.29 | 1.21 |
| PDM 2 bits shaped | 27.48 | 0.40 |
| PDM 3 bits shaped | 27.35 | 0.27 |
| PDM 4 bits shaped | 27.32 | 0.24 |
| PDM 5 bits shaped | 27.31 | 0.23 |

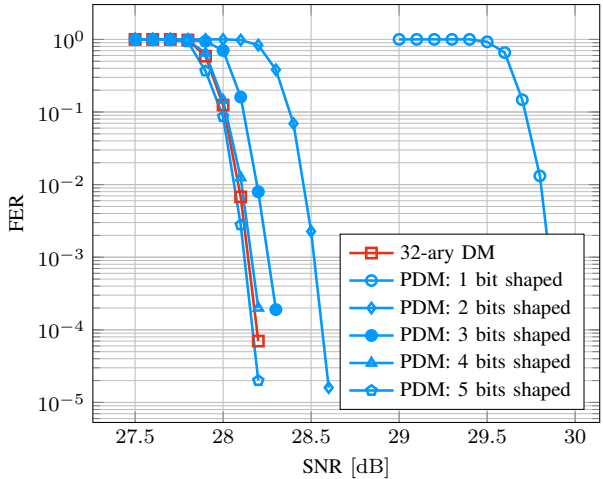

Fig. 3. Performance comparison of the proposed PDM for 64-ASK and a target SE of 4.5 bpcu and different number of shaped bits.

We observe that the gap to capacity to achieve an SE of 4.5 bpcu is very similar for 3, 4 and 5 shaped bit-levels, larger gaps can be observed when only 1 or 2 bit-levels are shaped.

We compare the asymptotic results to a finite length scenario with a rate 9/10 LDPC code (corresponding to $\gamma = 0.4$) from the DVB-S2 standard with a blocklength of 64 800 bits and a corresponding DM output length of 10 800 symbols. We observe that the performance of the 32-ary DM does not improve upon PDM with four or five shaped bit-levels in contrast to the asymptotic results of Table II. This is because of the larger output alphabet and the slower convergence of the DM rate to its asymptotic limit. One hundred iterations are used for the belief propagation (BP) decoding.

## IV. PDM FOR PARALLEL CHANNELS

We now consider $L$ parallel channels of the same form as in (4)

$$Y_\ell = h_\ell X_\ell + Z_\ell, \quad \ell = 1, 2, \ldots, L. \tag{18}$$

The noise terms $Z_\ell$ are zero mean Gaussian with unit variance. The coefficients $h_\ell$ model the channel gains and we assume that both the receiver and transmitter have full channel state information, i.e., they both know the channel gains $h_\ell$ and the noise variance.

### A. Waterfilling Benchmark

The transmitter has an average power budget $P$, i.e., the inputs are subject to the sum-power constraint

$$\frac{1}{L} \sum_{\ell=1}^{L} \mathrm{E}\left[X_\ell^2\right] \leq P. \tag{19}$$

The average SE

$$\frac{1}{L} \sum_{\ell=1}^{L} \frac{1}{2} \log_2(1 + h_\ell^2 P_\ell) \tag{20}$$

is achievable with the channel inputs $X_\ell$ being independent zero mean Gaussian with variance $P_\ell$. The average SE is maximized by waterfilling, i.e.,

$$P_\ell^* = \left[\frac{1}{\lambda} - \frac{1}{h_\ell^2}\right]^+, \quad \lambda \colon \frac{1}{L} \sum_{\ell=1}^{L} P_\ell^* = P. \tag{21}$$

Suppose that $P_\ell^*$ is positive. The SE allocated to channel $\ell$ is then $C_\ell = \frac{1}{2} \log_2(h_\ell^2/\lambda)$ and we have

$$\mathsf{C}_{\mathrm{WF}}(P) = \frac{1}{L} \sum_{\ell=1}^{L} C_\ell = \frac{1}{L} \sum_{\ell=1}^{L} \frac{1}{2} \log_2 \frac{h_\ell^2}{\lambda}. \tag{22}$$

The function $\mathsf{C}_{\mathrm{WF}}(P)$ is the maximum achievable SE under the sum power constraint $P$ and it serves in the following as our benchmark. For discrete inputs, the power allocation follows the mercury-waterfilling principle [20]. In the following, we develop a heuristic that uses PDM and operates closely to $\mathsf{C}_{\mathrm{WF}}(P)$.

### B. Bit-Loading Strategy

Since the per-channel SEs can differ by several bits, we need to support several constellations in parallel. This is for instance important for digital subscriber line (DSL) systems, where some good channels may support up to 32 768-QAM [21], whereas the majority needs to be operated with smaller modulation formats. Next, we have to decide which constellation size is used for which channel, an approach known as bit-loading. We employ the following heuristic: We calculate the waterfilling solution for the given channel coefficients and obtain the optimal rate assignment $C_\ell, \ell = 1, \ldots, L$ from (22). Then, we use Ungerböck's rule-of-thumb [22] to choose a constellation size $M_\ell = 2^{m_\ell}$ for channel $\ell$ such that $m_\ell \approx C_\ell + 1$. This avoids a reduced SE because of too small constellation sizes. We assume the largest constellation size is $2^m$, i.e., $m = \max_\ell m_\ell$. Further, the smallest constellation size is $2^2$-ASK for the ease of exposure. An extension to channels using binary phase-shift keying (BPSK) is straightforward.

### C. PAS for Parallel Channels

PAS can be combined with parallel channels as illustrated in Fig. 4a. A DM device transforms data bits into a sequence of amplitudes for each channel, which are then combined with sign bits originating from a common encoding device. In its simplest form, this DM device internally uses individual DMs, each with its output alphabet size matched to the corresponding constellation size.

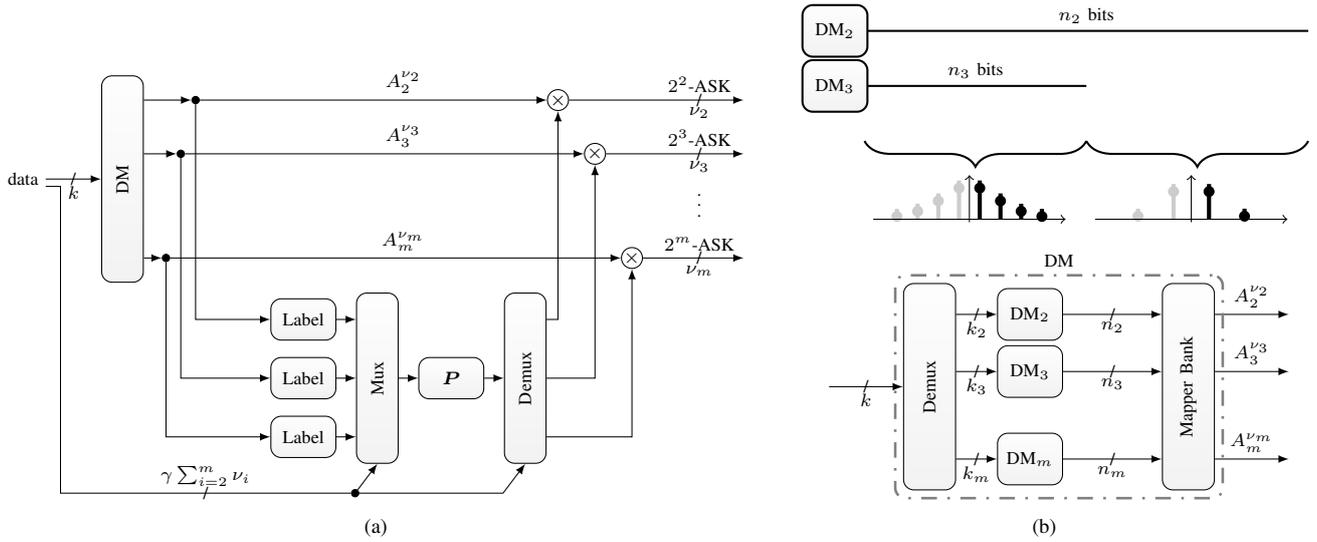

Fig. 4. (a) Illustration of PAS for $L = \sum_{i=2}^{m} \nu_i$ parallel channels. Note that the power control can still be applied individually. (b) Generating two Gaussian-like amplitude distributions for 4-ASK and 8-ASK simultaneously by reusing the DM of bit-level (c) The PDM architecture for parallel PAS transforms $k$ data bits into $m$ amplitude sequences of lengths $\nu_2, \ldots, \nu_m$. Internally, PDM for parallel PAS uses $m-1$ binary component DMs, where $2^m$-ASK is the largest supported constellation. The output lengths $n_2, \ldots, n_m$ of the component DMs are given by (24). The input lengths fulfill $\sum_{i=2}^{m} k_i = k$.

### D. PDM for Parallel Channels

PDM allows to jointly generate a length $L$ amplitude sequence with different constellation sizes. For example, suppose we have $L = \nu_2 + \nu_3$ possibly different channels where $\nu_2$ channels use 4-ASK and $\nu_3$ channels use 8-ASK. The PDM needs one binary DM for 4-ASK and two binary DMs for 8-ASK. As illustrated in the top part of Fig. 4b, the idea is now to use for the first amplitude bit-level $B_2$ of 4-ASK and 8-ASK a single binary DM with output length $n_2 = \nu_2 + \nu_3$ and to generate the second amplitude bit-level $B_3$ for 8-ASK by a second binary DM with output length $n_3 = \nu_3$. This approach allows the DMs to operate over a longer blocklength, causing the DM rate to reach its asymptotic limit faster. The illustration in the bottom part of Fig. 4b shows this scheme.

### E. Parametrization

We state how the system has to be parameterized to operate at a given SE. For the considered case we assume $\nu_i$ channel uses of a $2^i$-ASK constellation for $i = 2, \ldots, m$ within one FEC frame. The blocklength of the FEC code is

$$n_c = L + \sum_{i=2}^{m} n_i \qquad (23)$$

where we have $L = \sum_{i=2}^{m} \nu_i$ and the parameters $n_i$ denote the DM output lengths

$$n_i = \sum_{\ell=1}^{L} \mathbb{1}(m_\ell \geq i) \cdot \nu_i, \quad i = 2, 3, \ldots, m. \qquad (24)$$

The function $\mathbb{1}(\cdot)$ is the indicator function, which evaluates to one if its argument is true and zero otherwise. The corre-

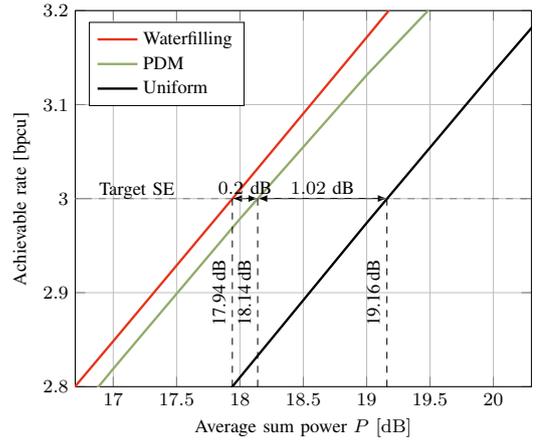

Fig. 5. Achievable rates of the considered example.

sponding DM input lengths are $k_2, k_3, \ldots, k_m$. The average SE of the overall system is now

$$R_{\text{tx}} = \frac{\sum_{i=2}^{m} k_i}{\sum_{i=2}^{m} \nu_i} + \gamma. \qquad (25)$$

and converges to $(\sum_{i=2}^{m} \mathrm{H}(B_i^{\text{dm}}) n_i)/(\sum_{i=2}^{m} \nu_i) + \gamma$ for large $L$. The formulas (7) and (8) generalize to

$$R_c = \frac{\sum_{i=2}^{m}(i-1+\gamma)\nu_i}{\sum_{i=2}^{m} \nu_i \cdot i} \qquad (26)$$

$$\gamma = 1 - (1 - R_c)\frac{\sum_{i=2}^{m} \nu_i \cdot i}{\sum_{i=2}^{m} \nu_i}. \qquad (27)$$

### F. Simulation Results

To evaluate the performance of parallel PAS with PDM, we employ the following example of three different constellation sizes which are used equally often. The coefficients are chosen such that the channel quality varies significantly (over a range of 12 dB) and requires three different modulation formats. The waterfilling solution (22) for a target SE of 3.0 bpcu yields the following rate allocation

$$Y_1 = 2.0 \cdot X_1 + Z_1, \quad C_1 = 4.0$$
$$Y_2 = 1.0 \cdot X_2 + Z_2, \quad C_2 = 3.0$$
$$Y_3 = 0.5 \cdot X_3 + Z_3, \quad C_3 = 2.0$$

which is achieved for an average sum-power of 17.94 dB. We select constellation sizes of $2^{m_1} = 32$, $2^{m_2} = 16$ and $2^{m_3} = 8$ points according to our bit-loading strategy of Sec. IV-B. The achievable rates are plotted over the average sum-power in Fig. 5. Our proposed heuristic scheme exhibits a gap of 0.2 dB to the waterfilling benchmark of (22) for the target SE of 3.0 bpcu. The uniform reference curve is shown in black and has a gap of 1.22 dB to the waterfilling solution. The employed bit distributions are summarized in Table III and have been chosen as the solution to the following heuristic optimization problem for $R_{\text{dm}} = 3.0 - \gamma$ with $\gamma = 1/3$:

$$\min_{P_{B_2^{\text{dm}}}, \ldots, P_{B_m^{\text{dm}}}} \sum_{\ell=1}^{3} \frac{1}{h_\ell^2} \operatorname{E}\left[X_\ell^2\right] \text{ s.t. } \frac{1}{L} \sum_{i=2}^{5} \operatorname{H}(B_i^{\text{dm}}) n_i = R_{\text{dm}}. \tag{28}$$

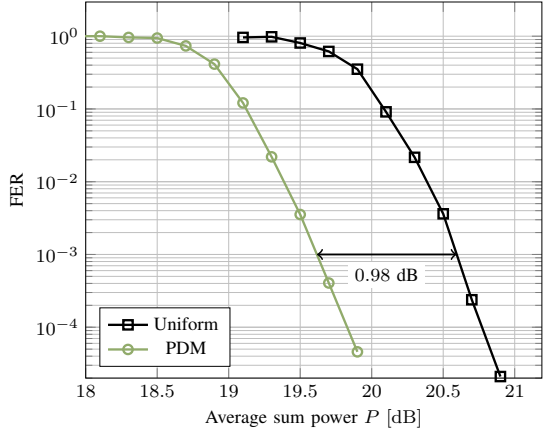

Fig. 6. Coded performance comparison of PDM and uniform scheme for parallel channels (LDPC code with block length 3600 bits).

In Fig. 6, we consider the same scenario with finite length LDPC codes from the 5G standard [3] (basegraph BG1). The uniform reference uses a rate $R_c = 3/4$, while the shaped case has a rate $R_c = 5/6$ code ($\gamma = 1/3$). In both cases the number of transmitted bits is 3600. The asymptotic gains are also reflected in the coded results. We perform 100 BP iterations.

TABLE III
PDM PROPERTIES FOR THE CONSIDERED EXAMPLE.

| $\text{DM}_i$ | $\nu_i$ | $n_i$ | $P_{B_i^{\text{dm}}}(0)$ | $\operatorname{H}(B_i^{\text{dm}})$ |
|---|---|---|---|---|
| 2 | 300 | 900 | 0.1995 | 0.7208 |
| 3 | 300 | 900 | 0.3736 | 0.9534 |
| 4 | 300 | 600 | 0.4408 | 0.9898 |
| 5 | 300 | 300 | 0.4709 | 0.9976 |

## V. CONCLUSION

We proposed product distribution matching (PDM), an architecture that uses binary DMs in parallel. This parallelization enables high-throughput implementations of DMs and the binary component DMs of PDM reduce complexity. We also proposed PDM for parallel PAS, which enables an operation close to the waterfilling limit of multi-carrier transmission schemes such as OFDM and improves over a uniform reference by 1 dB in a representative scenario.